\documentclass[twocolumn,showpacs,preprintnumbers,amsmath,amssymb]{revtex4}
\usepackage{graphicx}% Include figure files
\usepackage{dcolumn}% Align table columns on decimal point
\usepackage{bm}% bold math

\begin{document}

%\preprint{APS/123-QED}
\title{Extensive analytical and numerical investigation of the kinetic and stochastic Cantor set}%

\author{M. K. Hassan $^{1}$, M. Z. Hassan$^{2}$ and N. I. Pavel$^{1}$
}%
%\email[REVTeX Support: ]{revtex@aps.org}
%\affiliation{1 Research Road, Ridge, NY 11961}
\date{\today}%

\affiliation{
$1$ University of Dhaka, Department of Physics, Theoretical Physics Group, Dhaka 1000, Bangladesh \\
$2$ Information and Communication Technology Cell, Bangladesh Atomic Energy Commission, Dhaka 1000, Bangladesh 
}

\begin{abstract}
We investigate, both analytically and numerically, the kinetic and stochastic counterpart of the triadic
Cantor set. The generator that divides an interval either into three equal pieces or into three pieces randomly
and remove the middle third is applied to only one interval, picked with probability proportional to its size, 
at each generation step in the kinetic and stochastic Cantor set respectively. We show that the fractal dimension of 
the kinetic Cantor set coincides with that of its classical counterpart despite the apparent differences in the spatial 
distribution of the intervals. For the stochastic Cantor set, however, we find that the resulting set has fractal 
dimension $d_f=0.56155$ which is less than its classical value $d_f={{\ln 2}\over{\ln 3}}$. Nonetheless, in all three 
cases we show that the sum of the $d_f$th power, $d_f$ being the fractal dimension of the respective set, of all the 
intervals at all time is equal to one or the size of the initiator $[0,1]$ regardless of whether it is recursive, kinetic 
or stochastic Cantor set. Besides, we propose exact algorithms for both the variants which can capture the 
complete dynamics described by the rate equation used to solve the respective model analytically. 
The perfect agreement between our analytical and numerical simulation is a clear testament to that. 
\end{abstract}


\maketitle

\section{Introduction}

The history of describing natural objects by geometry is as old as the history of science itself.
Perhaps, the oldest of our pedagogical understanding about the properties of physical objects is geometry. 
Thanks to the Greek philosopher Euclid who is in fact the principal architect for laying the early foundation of 
geometry which is now known as the Euclidean geometry. For centuries, it has been the only means of describing geometry 
of physical objects. However, the then scientists also realized that the nature is not restricted to Euclidean space only; 
instead, most of the natural objects we see around us are so complex in shape that conventional Euclidean space is 
not sufficient to describe them. It was not until the work of Benoit B. Mandelbrot that dramatic progress 
was made. In 1975 Mandelbrot introduced the idea of fractal 
that has revolutionized the whole concept of geometry \cite{ref.mandelbrot1}. 
Prior to the inception of fractal, geometry remained one of the main branches of mathematics. However,
soon after its inception, it has 
attracted mathematicians, physicists, and engineers all alike and hence generated a widespread interest.
All credit goes to Mandelbrot for the way he presented the idea of fractal through his monumental book 
{\it The Fractal Geometry of nature} \cite{ref.mandelbrot2}. Indeed, he presented his book in an unusually 
inspiring way and since then it remained as the most favourite standard reference book for both beginners and 
researchers. Due to its wide interest, it has brought many seemingly unrelated subjects under one umbrella
and provided a tools to appreciate that there exists some kind of order even in
the seemingly complex and apparently disordered many natural geometric structures.

The importance of fractal and multifractal in nonlinear dynamics especially in chaos can hardly be exaggerated.
Their relation exist to such an extent that 
the impact of the ideas of chaos and fractals in
physics and other scientific disciplines in recent years
have been enormous. One common thing about both chaos and fractals is that much of their progress was possible due to 
high configuration computers and numerical simulations.  
Interestingly, the real progress of fractals and chaos both began during the 1970s.
 Most of the books or articles written on chaos are found to
invoke the idea of fractal since the plot of the fractal basins associated with chaos 
provides a qualitative idea of the extent of complications in the prediction of its future evolution
\cite{ref.rmp}. For this and because
of their common history often general people regards the two concepts synonymous. Note that the keywords in chaos
are non-linearity, unpredictability, sensitivity to initial conditions. On the other hand, the keywords in fractals 
are self-similarity and scale-invariance that is, the objects look the same on different
scales of observation.

The exact definition of fractal even after all these years is still elusive.
Mandelbrot himself was somehow reluctant to confine it within the 
boundaries of a mere definition. He, nevertheless proposed that fractal can be defined as 
a geometric object which is similar to itself on 
all scales. That is, if one zooms in on a fractal object it will look exactly similar to the original shape. 
Mandelbrot offered also a mathematical definition  ''a fractal is by definition a set for which 
the Hausdorff-Besicovitch dimension strictly exceeds the topological dimension". 
Sometimes fractal is also defined in terms of mass-length relation \cite{ref.vicsek}. That is,
if the mass $M$ of an object is related to its different possible size $L$ by the relation
\begin{equation}
\label{masslength}
M(L)\sim L^{d_f},
\end{equation}
where $d_f$ is less than the dimension of the space in which the object is embedded then the object is called 
fractal and it is quantified by its dimension $d_f$. 
%The mass-length relation given by 
%Eq. (\ref{masslength}) can be used to obtain yet another definition of fractal.
%An object which satisfies Eq. (\ref{masslength}) has density 
%\begin{equation}
%\rho(L)={{{\rm Mass\ of\ the \ object}}\over{{\rm Volume \ of \ the\ embedding\ space}}}\sim L^{d_f-d_E},
%\end{equation}
%where $d_E$ is the dimension of the embedding space and it is always Euclidean in character.
%That is, an object is a fractal if it becomes less dense at larger length scale since its dimension $d_f$ is less than 
%the dimension of the embedding space $d_E$. This scale dependence density is in 
%fact a quantitative measure of the degree of ramification or stringyness of fractal objects. 

The Hausdorff-Besicovitch dimension actually plays the pivotal role for a rigorous definition of fractal
and for providing a procedure to find the fractal dimension. Consider that the measure $M_d$ is the size of the
set of points in space of the object. We can quantify the size of the measure $M_d$ by using $d$-dimensional
hypercube of linear size $\delta$ as a yardstick to find the number $N(\delta)$ needed to cover $M_d$ \cite{ref.feder}. 
That is, we can cover the object to form the measure 
\begin{equation}
\label{eq:measure}
M_d=\sum \delta^d=N(\delta)\delta^d,
\end{equation}
where $\delta^d$ is the test function. The number $N(\delta)$
will obviously gets smaller as the size of the yardtick $\delta$ gets larger. It can be easily shown that $N(\delta)$
satisfies the following generalized relation
\begin{equation}
\label{eq:gdim5}
N(\delta/n)=n^{d_E} N(\delta),
\end{equation}
if $M_d$ describes an Euclidean object. 
The above relation for $N$ satisfies the property of a homogeneous function $N(\lambda x)=\lambda^pN(x)$ if
one chooses $\lambda=n^{-1}$, $x=\delta$ and $p=-d_E$.
One can explicitly prove that only power-law solution can satisfy Eq.
(\ref{eq:gdim5}) \cite{ref.newman}. Indeed, it is easy to check that $N(\delta)\sim \delta^{-d_E}$ can solve 
Eq. (\ref{eq:gdim5}).
In reality there exist another class of objects which also exhibits a similar power-law relation 
between the number $N$ and the size of the yardstick $\delta$ but with a non-integer exponent $d_f$ instead of $d_E$ and hence
one can generalize the relation as 
\begin{equation}
\label{eq:dim6}
N(\delta)\sim \delta^{-D},
\end{equation}
where $D$ is called the Hausdorff-Besicovitch (H-B) dimension \cite{ref.feder}. Using this relation in Eq. (\ref{eq:measure}) we find that the
value of $M_d$ assumes a finite constant value only if $d=D$ otherwise the measure $M_d$ is either zero if $d>D$  
or infinity if $d<D$ in the limit $\delta \rightarrow 0$. The H-B dimension therefore
is the critical dimension $d=D$ for which the measure $M_d$ neither vanishes
nor diverges as $\delta \rightarrow 0$. So, an object is fractal if the 
H-B dimension $D=d_f$ a non-integer value.

The best known text example of fractal is the triadic Cantor set. 
However, this classical Cantor set lacks in 
two ways from the natural fractals. Firstly, it does not appear through evolution in time
although fractals in nature do so. Secondly, it is not governed by any sort of randomness 
throughout its construction. On the other hand, natural fractals always occur through some kind of evolution 
accompanied by some randomness.
In this article, we therefore, investigate two interesting variants of the classical Cantor set, 
the kinetic and the stochastic Cantor set, in which time and randomness are incorporated in a logical progression. 
The kinetic Cantor
set differs from the classical Cantor set in the sense that at each step only one of the available
interval is picked, with probability proportional to its size, for dividing it into three equal pieces and remove the
middle third. On the other hand, in the stochastic Cantor set the definition of the generator is modified in the
sense that it
divides an interval into three pieces randomly and remove the middle third. However, the modified generator is
applied sequentially to only one interval which is also picked according to its size at each generation step like 
in the kinetic Cantor set. The adventage of applying the generator sequentially is that we can invoke time as one 
of the parameter in the problem. On the other hand modifying the generator in the stochastic Cantor set
incorporate randomness into the problem as it evolves. 
In fact, it is well understood that in our world almost nothing is stationary
or strictly deterministic. Every natural objects are seemingly complex in character though they are governed
by simple rules since simple rules when repeated over and over again can appear mighty complex. 

The Cantor set provides us with a wealth of interesting properties which are
taught in advanced undergraduate and graduate studies in discrete mathematics course.
Apart from its pedagogical importance the Cantor set problem has also been of theoretical and practical interest. 
For instance, Sears {\it et al} has shown that a nonlinear system that supports solitons can be driven to generate 
Cantor set \cite{ref.sears}. In another case, it has been found that the electromagnetic wave is strongly enhanced and 
localized in the cavity of the Cantor set near the resonant frequency \cite{ref.hatano}. 
Krapivsky and Redner shown that the probability distribution in a random walk
with a shrinking steps is a Cantor set \cite{ref.redner}. Recently, Esaki {\it et al} studied propagation 
of waves through Cantor set media and found some very interesting results such as complete reflection or complete
transmission including scaling property of the transmission co-efficients \cite{ref.esaki}.

The remainder of this article is organized as follows. In section II, titled 'recursive Cantor set'
we discussed the well-known triadic Cantor set. In section III, 
we proposed the kinetic counterpart of the triadic Cantor set and solved it exactly to obtain the fractal dimension
and various other properties. We also proposed exact algorithm for the kinetic Cantor set 
to solve the model by numerical simulation. In section IV, we investigated the stochastic counterpart
of the triadic Cantor set and once again we proposed its exact algorithm to solve it by numerical simulation.
In section V, we summarized our results.

\section{Recursive Cantor set (RCS)}

The novelty in the definition of the triadic Cantor set is its simplicity. The notion of fractal
and its inherent character, self-similarity, is almost always introduced to the beginner through this example.
The triadic Cantor set can be defined as follows.
It starts with an initiator of unit interval $[0,1]$. The generator then divides it into three equal parts 
and deletes the middle third leaving behind two sub-intervals, each with size one-third of the original interval 
such as $[0,{{1}\over{3}}]$ and $[{{2}\over{3}},1]$. In the next step, the generator is again applied to each of
the two sub-intervals that divides them into three equal parts of size ${{1}\over{9}}$ and remove the middle third from both. 
The process is then continued by applying the generator on the remaining intervals recursively {\it ad infinitum} and
hence we call it recursive Cantor set (RCS).
Like its definition, finding the fractal dimension of the RCS problem is also trivially simple. According to the construction 
of the RCS process there are $N=2^n$ intervals in the $n$th generation each of size $\delta=3^{-n}$ and hence
it is also the mean interval size. The most convenient yard-stick in the
$n$th step, therefore, is the mean interval size $\delta=3^{-n}$. The generation
number $n$ can be written as
\begin{equation}
n=-{{\ln \delta}\over{\ln 3}}.
\end{equation} 
Using it in $N=2^n$ we find that the number $N$ falls off following power-law
 against mean interval size $\delta$ i.e.,
\begin{equation}
\label{eq:fractal}
N(\delta)\sim \delta^{-d_f},
\end{equation}
with $d_f={{\ln 2}\over{\ln 3}}$. Since the exponent $d_f$ of the above relation is non-integer and at the same time 
it is less than the dimension of the space $d=1$ where the set is embedded, it is the fractal dimension of the resulting
triadic Cantor set. Note that the set does not fill up the unit interval 
by uniform distribution of zero-dimensional points to describe a line rather, it fills up the unit interval 
by zero-dimensional points (also known as the Cantor dust) in such a special way that 
it possess exact self-similarity.

\section{Kinetic Cantor set (KCS)}

One may wonder what if we divide only one interval at each step instead of dividing every available intervals  
as done in the RCS problem? 
Clearly, the spatial distribution of intervals along the line will be very different from the one created by the 
RCS problem. 
But, then the question is: Will the number $N$ needed to cover the set by an yardstick, say of size $\delta$, still
exhibit power-law against $\delta$? If yes, will the exponent {\it vis-a-vis} the 
fractal dimension be the same as for the RCS? To find a definite answer to these questions, 
the same generator that
divides an interval into three equal pieces and remove the middle third in the RCS problem is applied to only 
one of the available intervals instead of applying it to each of the available intervals at each step. But 
after step one and beyond the system will have intervals of different sizes and hence it raises further
question: How do we choose one interval when the system has more than one interval of different sizes? We choose
the case whereby an interval is picked with probability proportional to their respective sizes
as it appears to be the most generic case. One advantage of modifying the RCS problem in this way is that we can use the
rate equation approach to solve it analytically. Since time becomes one parameter of the problem  
we call it kinetic Cantor set (KCS).

The KCS problem can be defined as follows. Like RCS it may also starts with an initiator of unit interval 
$[0,1]$. In the first step the generator therefore divides the initiator into three sub-intervals of equal size
and remove the middle third. The two newly created intervals are labelled as $1$ and 
$2$ starting from the left end of the two surviving intervals. In the next
step we generate a random number $R$ from the open interval $(0,1)$ and find which of the two 
subintervals $1$ and $2$ contains this number $R$ in order to ensure that intervals are picked according to their size. 
If $R$ is found within the interval $[0,{{1}\over{3}}]$ then we pick interval $1$, if it is found
within $[{{2}\over{3}},1]$ we pick interval $2$. Say, the interval $1$ contain $R$ and hence we pick interval $1$.
The generator then divides it into three equal pieces and remove the middle third. The left end of the two newly
created interval is then labelled with its parent label $1$ and the intervals on its right is labelled
as $3$. In any case, time is increased by one unit even if $R$ is found within
the interval that has been removed. Note that between two successive generation steps the time unit may increase
by several units since each time an attempt in picking an interval is unsuccessful the time is increased by one
unit. One therefore cannot write a straightforward relation between time $t$ 
and generation step $j$ although we will explore later that there do exist a non-trivial relation between $N=(j+1)$ and time
$t$. The $j$th step therefore starts with $j$ number of intervals 
labelled as $1,2,..., j$ and at the end of the $j$th step the system will have $j+1$ intervals whose sizes can be 
denoted as $x_1, x_2,...,x_{j+1}$. 
The basic algorithm of the $j$th step which starts with $j$ number of intervals can be described as follows. 
\begin{itemize}
\item [(a)] Generate a random number $R$ from the open interval $(0,1)$.
\item [(b)] Check which of the $1,2,...,j$ intervals contains the random number $R$. Say, the interval that contain $R$ 
is labelled as $m$ and hence pick the interval $m$.  Else, if none of the $j$ intervals contain $R$ then increase time 
by one unit and go to step $(a)$.
\item [(c)] Apply the generator to the sub-interval $m$ to divide
 it into three equal pieces and remove the middle third.
\item[(d)] Label two newly created intervals starting from the left end which is labelled with its parents label $m$ and 
the interval on the right is labelled with a new number $j+1$ that has not been already used. 
\item[(e)] Increase time by one unit.
\item[(f)] Repeat the steps $(a)$-$(e)$ {\it ad infinitum}.
\end{itemize}
The expression for the mean interval size
after the $j$th step therefore is $\delta=\sum_i^{j+1} x_i/(j+1)$ and the corresponding time $t$ can be obtained from
the counter used for time in the algorithm.

\begin{figure}
\includegraphics[width=8.50cm,height=4.5cm,clip=true]{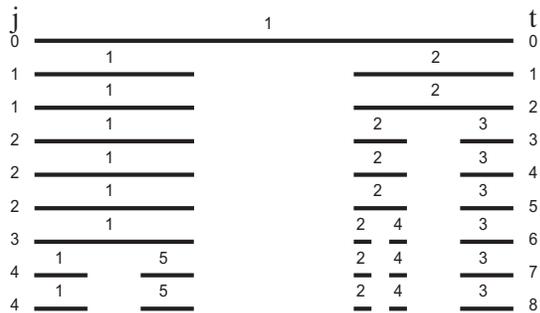}
\caption{Schematic illustration of the construction of the kinetic Cantor set.
On the left the numbers below $j$ indicates generation steps which is related to number of intervals via
$N=j+1$. On the right the numbers below $t$ indicates increase in time which is related to $N$ via a non-trivial
relation that we will explore.
}
\label{fig1}
\end{figure}

To solve the KCS problem analytically we use the rate equation approach.
The state of the system at any time $t$ can be
characterized fully by the interval size distribution function $C(x,t)$ which is defined in such a way
that $C(x,t)dx$ is the number of intervals in the size range $x$ and $x+dx$ 
at time $t$. The evolution of the distribution function $C(x,t)$ can then be described by the following master equation
\begin{eqnarray}
\label{eq:ternery}
{{\partial C(x,t)}\over{\partial t}} & = & -C (x,t)\int_0^xdy\int_0^{x-y}F(y,z|x)dz   
\nonumber \\ &+& 2\int_x^\infty dyC(y,t)\int_0^{y-x}F(x,z|y)dz,
\end{eqnarray}
where the kernel $F(x,y|z)$ describes the rate and rules how a parent interval of size $z$ is divided into three 
smaller intervals of size $x$, $y$ and $(z-x-y)$. The first term on the right hand side of the above rate equation
describes the loss of interval of size $x$ due to breakup of an interval of size $x$ while
the second term describes the gain of interval of size $x$ due to breakup of an 
interval of size $y>x$ into three smaller pieces so that at least one of the three smaller intervals is of size $x$.
The factor $2$ in the gain term guarantees that only two of the three intervals are kept
and one is removed which is exactly the definition of the Cantor set. 
Note that if the factor $2$ in Eq. (\ref{eq:ternery}) is replaced by the factor $3$, then the resulting equation 
describes the kinetics of ternary fragmentation process.
The ternary fragmentation equation was first proposed and solved exactly by Ziff \cite{ref.ziff}. In order to mimic 
the generator that picks an intervals according to size of the available intervals
and divide it into three equal pieces we choose 
\begin{equation}
\label{eq:kernel}
F(x,y|x+y+z)=(x+y+z)\delta(x-y)\delta(y-z).
\end{equation}
where the two delta functions ensures that the three intervals produced by the generator are equal in size.
Substituting this kernel into Eq. (\ref{eq:ternery}) we get
\begin{equation}
\label{eq:ternery_1}
{{\partial C(x,t)}\over{\partial t}}=-{{x}\over{3}}C(x,t)+6xC(3x,t).
\end{equation}
One can solve the above equation exactly to find the solution for $C(x,t)$ using the Laplace transformation.

We are not really interested in the solution for the distribution function $C(x,t)$ rather we are interested in finding
solution of its $n$th moment defined as
\begin{equation}
\label{eq:moment}
M_n(t)=\int_0^\infty x^nC(x,t)dx,
\end{equation}
as it can provide almost all the information that we intend to find.
Moreover, we find it more convenient to analyze the moment of the distribution function than the function itself.
Incorporating the definition of the $n$th moment in Eq. (\ref{eq:ternery_1}) yield the following rate equation for the
$n$th moment
\begin{equation}
\label{eq:moment_c}
{{M_n(t)}\over{dt}}=\Big [ {{1}\over{3}}-{{2}\over{3^{n+1}}} \Big ]M_{n+1}(t).
\end{equation}
It is interesting to note that the distribution function $C(x,t)$ is not a directly observable quantity but its 
various moments are, 
for instance, $M_0(t)=N(t)$ is the number of available intervals at time $t$, $M_1(t)=L$ is the sum of the sizes
of all the available intervals at time $t$ etc. This clearly justifies the reason behind focusing on finding the solution 
for the various moments than the distribution function itself. For consistency check, let us consider a case whereby none of
intervals are removed after generator divides an interval into three equal pieces. That is, the sum of the sizes 
of all the intervals at all time would be equal to the size of the initiator. The corresponding rate equation for $M_n(t)$ 
which can be obtained from Eq. (\ref{eq:moment_c}) upon replacing the factor $2$ by $3$ from which one can easily
find that 
\begin{equation}
{{dM_1}\over{dt}}=0,
\end{equation}
and hence $M_1(t)=L$ is indeed independent of time which is infact equal to the size of the initiator.

In order to obtain a complete solution for the $n$th moment of Eq. (\ref{eq:moment_c}) we assume that there exists a value
$n=n^*$ so that $M_{n^*}$ remains independent of time or a conserved quantity in the long time limit.
Indeed, a closer look into the rate equation for the $n$th moment
immediately implies that we can obtain the value of $n^*$ by applying the steady-state 
condition
\begin{equation}
\label{eq:moment_d}
\lim_{t\longrightarrow \infty}{{dM_n(t)}\over{dt}}=0,
\end{equation}
in Eq. (\ref{eq:moment_c}) and hence obtain the following equation
\begin{equation}
{{1}\over{3}}-{{2}\over{3^{n^*+1}}}=0.
\end{equation}
Solving it for $n^*$ we immediately find that $n^*={{\ln 2}\over{\ln 3}}$ which implies that 
$M_{{{\ln 2}\over{\ln 3}}}$ is a conserved quantity.
One of the property of fractal is that it must obey scaling or self-similarity. As we are expecting that the KCS problem 
like its cousin RCS problem will also generate fractal in the long time limit. It is therefore reasonable to anticipate
that the solution of Eq. (\ref{eq:ternery_1}) for general $n$ will exhibit scaling. 
Existence of scaling means that the various moments of $C(x,t)$
should have power-law relation with time and hence we
can write a tentative solution of Eq. (\ref{eq:moment_c}) as below  
\begin{equation}
M_n(t) \sim A(n)t^{\alpha(n)}.
\end{equation}
If we insist that it must obey the conservation law, $M_{{{\ln 2}\over{\ln 3}}}=const.$ then we must have 
$\alpha(\ln 2/\ln 3)=0$. Substituting this in Eq. (\ref{eq:moment_c}) we obtain the following recursion relation
\begin{equation}
\alpha(n+1)=\alpha(n)-1.
\end{equation}
Iterating it subject to the condition that $\alpha(\ln 2/\ln 3)=0$ gives
\begin{equation}
\alpha(n)=-(n-{{\ln 2}\over{\ln 3}}).
\end{equation}
We therefore now have an explicit asymptotic solution for the $n$th moment  
\begin{equation}
\label{eq:moment_e}
M_n(t) \sim t^{\Big(n-{{\ln 2}\over{\ln 3}} \Big )Z}, \hspace{0.35cm} {\rm where}  \hspace{0.35cm}
Z=-1.
\end{equation}

According to Eq. (\ref{eq:moment_e}) we find that the number of intervals $N(t)=M_0(t)$ grows as
\begin{equation}
\label{eq:number}
N(t)\sim t^{{{\ln 2}\over{\ln 3}}},
\end{equation}
and the mass or the sum of all the intervals size decreases against time as
\begin{equation}
\label{eq:mass}
M(t)\sim t^{-0.369}.
\end{equation}
The solutions for $M_0(t)$ and $M_1(t)$ can provide us with the information how the mean interval size 
$\delta(t)=M_1(t)/M_0(t)$ decay and find that 
\begin{equation}
\label{eq:mean}
\delta \sim t^{Z},
\end{equation}
where the kinetic exponent $Z=-1$. 
We now apply the idea of fractal analysis which is briefly described in the introduction. To this end 
we find it convenient to use typical or mean interval size $\delta(t)$ 
as the yard-stick to measure the resulting set since it will always 
give an integer $N$. 
This is equivalent to expressing the number $N$ in terms of $\delta$ and in doing so we 
find that $N(\delta)$ decreases following the same power-law as Eq. (\ref{eq:fractal}) including its exponent 
$d_f={{\ln 2}\over{\ln 3}}$. 
The H-B dimension of the resulting KCS problem 
therefore is $D={{\ln 2}\over{\ln 3}}$  which is exactly the same as its recursive counterpart
albeit the spatial distribution is very different.

\begin{figure}
\includegraphics[width=8.50cm,height=4.5cm,clip=true]{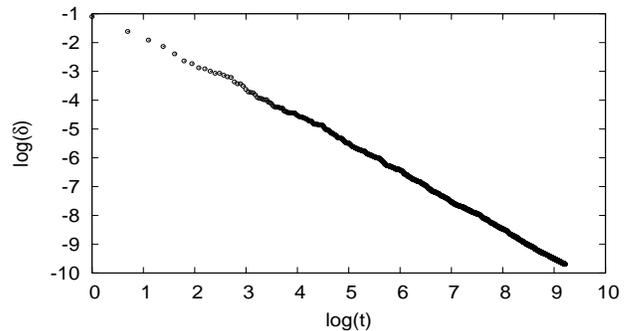}
\caption{The decrease in mean interval size $\delta$ against time $t$, as defined in the algorithm,
is drawn in the $\log-\log$ scale using numerical data and found that it satisfies $\delta\sim t^{-1}$
which is also predicted by analytical solution.
}
\label{fig2}
\end{figure}

\begin{figure}
\includegraphics[width=8.50cm,height=4.5cm,clip=true]{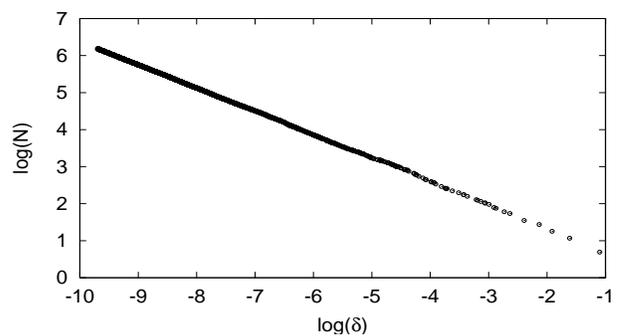}
\caption{Plots of $\log(N)$ vs $\log(\delta)$ is drawn using numerical data. The line has slope equal to
$d_f={{\ln 2}\over{\ln 3}}$ revealing $N\sim \delta^{-d_f}$ as predicted by the theory.
}
\label{fig3}
\end{figure}

\begin{figure}
\includegraphics[width=8.50cm,height=4.5cm,clip=true]{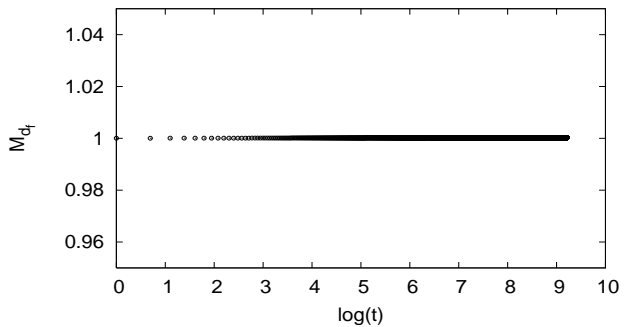}
\caption{The sum of the $d_f$th power of all the intervals $x_1^{d_f}+x_2^{d_f}+ ... + x_j^{d_f}=1$
regardless of time if $d_f={{\ln 2}\over{\ln 3}}$ which is equal to fractal dimension of the kinetic Cantor set. Its
analytical counterpart is the $d_f$th moment of the distribution function.}
\label{fig4}
\end{figure}

To verify our analytical results we performed Monte Carlo simulation based on the algorithm (a)-(f).
We first collected data for the mean intervals size $\delta=\sum_i^{j+1} x_i/(j+1)$ against time where
$x_i$s are the size of the intervals specified by their labels $i=1,2,....,j$. 
This data is used to draw $\log (\delta)$ vs $\log (t)$
in Fig. (2) and we clearly find a straight line with slope equal to $1$ revealing that the mean interval
size decreases following the same inverse power-law as predicted by Eq. (\ref{eq:mean}). 
Numerical data are also used to plot $\log (N)$ against $\log(\delta)$ and again we find a straight line but with a 
slope equal to ${{\ln 2}\over{\ln 3}}$ which is exactly as predicted by our theory (see Fig. (3)). It clearly proves
that the underlying mechanism described by the Eq. (\ref{eq:ternery_1}) has been captured by
the algorithm (a)-(f) we proposed for the KCS problem. One may think of another variant
of the kinetic Cantor set by removing one of the three intervals randomly each time generator divides an interval
into three equal intervals. Surprisingly, numerical data reveals that the value of $d_f$ still is the same regardless
of exactly which of the three intervals is removed each time an interval is divided into three equal intervals. 
Indeed, the rate equation for the distribution function $C(x,t)$ given
by Eq. (\ref{eq:ternery_1}) do not distinguish the three smaller intervals from one another.
Now, incorporating $d_f={{\ln 2}\over{\ln 3}}$ in Eq. (\ref{eq:moment_e}) we can conclude that the 
moment of order equal to fractal dimension $d_f$ is a conserved quantity. To verify this we collected data for 
the sum of the $d_f$th power of the size of all the intervals available at the generation step $j$  and 
found 
\begin{equation}
x_1^{d_f}+x_2^{d_f}+x_3^{d_f}+ ... \ ... \ + x_j^{d_f}=1,
\end{equation}
regardless of the value of $j$ {\it vis-a-vis} time $t$ which is equivalent to $M_{d_f}$. 
This is quite an extra-ordinary revelation. This result make us wonder if such conservation law also exist in the
case of traditional recursive Cantor set. Due to recursive nature of the RCS problem, after any generation
step $n$ all the $N=2^n$ intervals will have the same size $3^{-n}$. 
The sum of the $d_f$th power of all the intervals after the $n$th generation step therefore is 
\begin{equation}
2^n\Big (3^{-n}\Big)^{{{\ln 2}\over{\ln 3}}}=1,
\end{equation}
regardless of the value of $n$. We wonder what if we further modify the generator to create 
the stochastic fractal instead of random fractal. Will the relation that the sum of the $d_f$th power of all the intervals
at any given time be the same with that of at any other time?

\section{Stochastic Cantor set (SCS)}

The RCS problem hardly has any relevance to the fractal that appear in nature as it 
lacks at least in two ways from those that occur in
nature. For instance, it does not have any kinetics but most natural fractal appears through some kind of evolution
and it does not have any randomness but nature love to enjoy some degree of randomness to say the least. Though the KCS
problem appear through evolution but it still lacks in randomness. 
We therefore ask: What if we use a generator that divides an interval randomly into three smaller intervals instead of dividing
into three equal intervals? To find an answer to this question consider an algorithm for the stochastic Cantor 
set as described below. We start the process with an initiator of unit interval $[0,1]$ as before
but unlike the previous cases the generator here divides an interval randomly into three pieces. The algorithm
for the $j$th generation step that starts with $j$ number of intervals can be described as follows.  
\begin{itemize}
\item [(i)] Generate a random number $R$ from the open interval $(0,1)$.
\item [(ii)] Check which of the $1,2,...,j$ intervals contains $R$.
Increase time by one unit after every checking, starting from label $1$ then label $2$ and so on till an 
interval, say $[a,b]$ labelled as $k$, is found. Go to step $(i)$ if none of the available intervals contain $R$.
\item [(iii)] Apply the generator onto the interval $k$ to divide it randomly into three pieces.
For this we generate two random numbers from the open interval $(a,b)$, say $c$ and $d$ where say $c<d$, to divide the 
interval into $[a,c]$, $[c,d]$ and $[d,b]$ and delete the open interval $(c,d)$. 
\item [(v)] Update the logbook by labeling the left end of the two newly created interval $[a,c]$
with its parents label $k$ and right end of the two $[d,b]$ is labelled as $(j+1)$.  
\item[(vi)] Increase time by two units since two cuts are needed to divide an interval into three smaller intervals.
\item[(vii)] Repeat the steps $(i)$-$(vi)$ {\it ad infinitum}.
\end{itemize}

\begin{figure}
\includegraphics[width=8.50cm,height=4.5cm,clip=true]{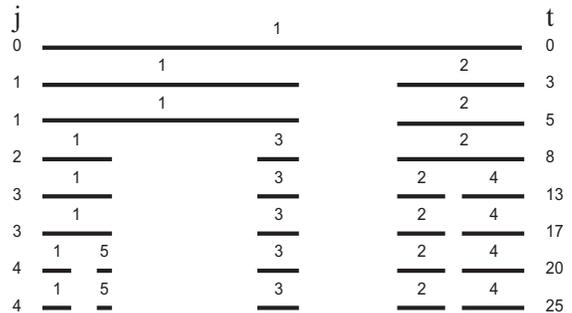}
\caption{Schematic illustration of the construction of stochastic Cantor set. The numbers on the top
of each intervals indicate their labels. The numbers on the left under $j$ indicates generation
steps and the numbers under $t$ on the right illustrate how time evolves. 
}
\label{fig5}
\end{figure}

\begin{figure}
\includegraphics[width=8.50cm,height=4.5cm,clip=true]{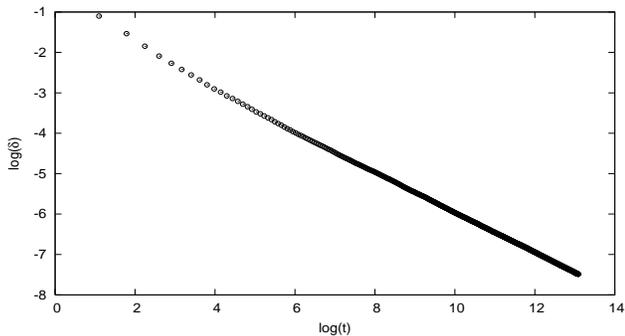}
\caption{Decrease in the mean interval size $\delta$ as a function of time $t$. The straight line in the $\log (\delta)$
vs $\log (t)$ plot with slope equal to ${{1}\over{2}}$ implies that $\delta\sim t^{-{{1}\over{2}}}$ and hence it
is in perfect agreement with analytical prediction. 
}
\label{fig6}
\end{figure}

The generalized master equation for the Cantor set given by Eq. (\ref{eq:ternery}) can describe the rules of the 
SCS problem stated in the algorithm (i)-(vii) if we choose 
\begin{equation} 
F(x,y|x+y+z) =1. 
\end{equation}
The master equation for the stochastic Cantor set then is 
\begin{equation}
\label{kn:SCS} 
{{\partial C(x,t)}\over{\partial t}}= -{{x^2}\over{2}}C(x,t)+2\int_x^\infty C(y,t)(y-x)dy.
\end{equation}
This is exactly the rate equation first proposed and solved analytically by Krapivsky 
in the context of random car parking problem and later by an alternative method
in the context of stochastic fractal \cite{ref.krapivsky, ref.hassan}. In this article, we however give an exact algorithm
that can capture the complete dynamics described by the above rate equation and verify the analytical results 
by numerical simulation.  
Following the method of Krapivsky and Ben-Naim we incorporate the definition of the $n$th moment in the rate
equation to give 
\begin{equation}
{{dM_n(t)}\over{dt}}=\Big (2{{\Gamma(n+1)\Gamma(2)}\over{\Gamma 
(n+3)}}-{{1}\over{2}}\Big ) M_{n+2}(t),
\end{equation}
and then solve it for $M_n(t)$. Following the same procedure as for the KCS problem we obtain the asymptotic 
solution for the $n$th moment
\begin{equation}
\label{eq:moment_scs}
M_n(t)\sim t^{(n-0.56155)z} \hspace{0.35cm} {\rm with}  \hspace{0.35cm}
z=-{{1}\over{2}},
\end{equation}
where the number $0.56155$ is the real positive root of the following quadratic equation
\begin{equation}
2{{\Gamma(n+1)\Gamma(2)}\over{\Gamma 
(n+3)}}={{1}\over{2}}.
\end{equation}
Note that once again we find that the exponent of the power-law relation for $M_n(t)$ is linear in $n$ and hence
the system must obey a simple scaling but only in the statistical sense. It is interesting to note that the moment of 
order $n=0.56155$, instead of $n=\ln 2/\ln 3$ in the KCS problem, is a consserved quantity. 
Using Eq. (\ref{eq:moment}) in the definition of the mean interval size we find that
\begin{equation}
\label{scs_mean}
\delta(t) \sim t^{-{{1}\over{2}}}.
\end{equation}
Once again we use it as the yard-stick and find that the number $N(\delta)$ 
needed to cover the resulting set scales as
\begin{equation}
\label{ndelta}
N(\delta) \sim \delta^{-d_f},
\end{equation}
where $d_f=0.56155$ is the fractal dimension of the stochastic Cantor set.
It reveals that the fractal dimension of the stochastic Cantor set is less than that of 
its recursive or kinetic Cantor set.

\begin{figure}
\includegraphics[width=8.50cm,height=4.5cm,clip=true]{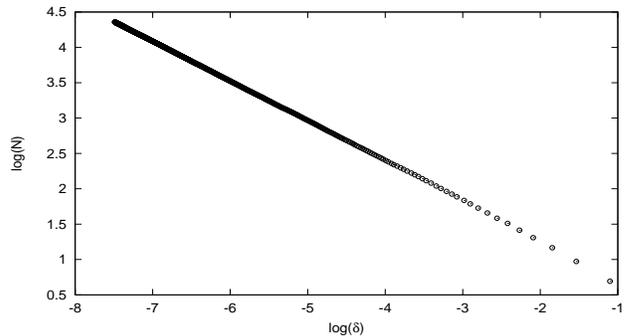}
\caption{Variation of the number of yard-sticks $N$ needed to cover the stochastic Cantor set is shown against the size
of the yard-stick $\delta$. Data collected from simulation is averaged over $5000$ independent realizations.
Slope of this straight line is found to be equal to $0.56155$ which means that $N(\delta)\sim \delta^{-0.56155}$.
}
\label{fig7}
\end{figure}

\begin{figure}
\includegraphics[width=8.50cm,height=4.5cm,clip=true]{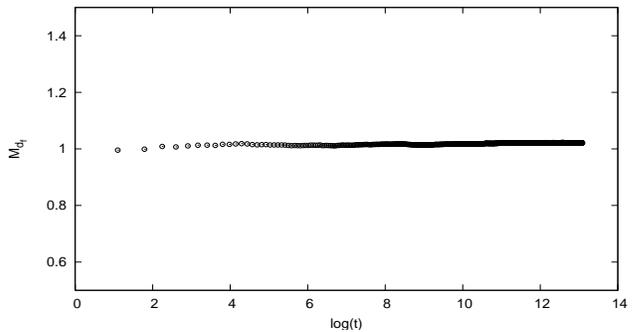}
\caption{The sum of the $d_f$th power of all the intervals $x_1^{d_f}+x_2^{d_f}+ ... + x_j^{d_f}=1$ 
regardless of time provided $d_f=0.56155$ which is the fractal dimension of the stochastic Cantor set. Data has been 
collected from the numerical simulation after averaging over $5000$ independent realizations. 
}
\label{fig8}
\end{figure}

To verify our analytical results we performed numerical simulation
based on the algorithm  $(i)-(vii)$.
Note that the definition of time in this algorithm is very much different from that of the kinetic Cantor set. 
Like in the KCS problem here too we collect data for the mean interval size $\delta$ against time. We then plot 
$\log (\delta)$ versus $\log (t)$ in Fig. (6) and find straight line with slope equal ${{1}\over{2}}$. It implies 
that the mean interval size decreases following exactly in the same fashion as predicted by Eq. (\ref{scs_mean}). 
We also collected data for $N$ against $\delta$. These data are used in Fig. (7) to show
how the number $N$ decreases with the yard-stick size $\delta$. 
A straight line in the logarithmic scale with slope equal to $0.56155$ clearly
implies that $N$ exhibits power-law relation $N\sim \delta^{-d_f}$ with exponent $d_f$ as predicted by Eq. (\ref{ndelta}).
Furthermore, in Fig. (8) we show that the sum of the $d_f$th power of all the intervals 
\begin{equation}
x_1^{0.56155}+x_2^{0.56155} + ... + ... + ...\ + x_j^{0.56155}=1
\end{equation}
regardless of the value of the generation step {\it vis-a-vis} the
time. From the analytical point of view this is equivalent to the moment $M_{0.56155}$ of the distribution 
function $C(x,t)$ which is indeed found to be independent of time. All these results clearly reveals that
analytical results are in perfect agreement with the numerical simulation.

\section{Summery}

To summarize, in this article we studied two interesting variant of the strictly self-similar triadic Cantor set. 
We solved the two models, the kinetic and stochastic Cantor set, both analytically using rate equation approach
and numerically based on the exact algorithms that we proposed for both the problems.
We found that the number of intervals and time are related via a generalized relation  
$N\sim t^{d_f}$ is true for both kinetic and stochastic Cantor set if we set 
$d_f={{\ln 2}\over{\ln 3}}$  for the KCS problem $d_f=0.56155$ for the SCS problem. On the other hand, we found that 
the mean interval size decreases with time following
power-law $\delta\sim t^z$ with exponent $z=1$ for the KCS problem and $Z={{1}\over{2}}$ for the SCS problem. We then took 
the respective mean interval size as the yard-stick to measure the resulting set
and found that the number $N$ needed to cover the set fall-off following power-law with exponent 
equal to their respective fractal dimension. We also found a generalized conservation law in the
sense that the sum of the $d_f$th power of all the available intervals at any time or the generation step
is equal to the size of the initiator. This is true for
recursive, kinetic and stochastic Cantor set. On the other hand, if we know the solution for the distribution function
$C(x,t)$, which can only be defined for kinetic and stochastic Cantor set, then the conservation law means 
that the $d_f$th moment of $C(x,t)$ remain independent of time. 
It is noteworthy to mention that such non-trivial conservation laws
was also found recently in the context of condensation-driven aggregation process \cite{ref.cda}. 
We can perhaps conclude that emergence of fractal in a given system
is always accompanied by some conservation laws which is ultimately responsible for fixing the various
scaling exponents including the fractal dimension.


\begin{thebibliography}{99}%

\bibitem{ref.mandelbrot1} Mandelbrot B B, Fractals, 1977 {\it Form, Chance, and Dimension} (Freeman, San Francisco)
\bibitem{ref.mandelbrot2} Mandelbrot B B, Fractals, 1982 {\it The Fractal Geometry of Nature} (Freeman, San Francisco)
\bibitem{ref.rmp} Aguirre J, Viana R L, Sanjuan M A F, 2009 Rev. Mod. Phys. {\bf 81} 333
%\bibitem{ref.peitgen} Peitgen H-O, Juergens H, and Saupe D, 2004 {\it Chaos and Fractals: New Frontiers of Science} 
%(Springer Verlag, New York)
\bibitem{ref.vicsek} Vicsek T, 1992 {\it Fractal Growth Phenomena}, 2nd ed. (World Scientific, Singapore)
\bibitem{ref.feder} Feder J, 1988 {\it Fractals} (Plenum Press, New York)
\bibitem{ref.newman} Newman M E J, 2005 Contemporary Physics, {\bf 46} 323
\bibitem{ref.sears}  Sears S, Soljacic M, Segev M, Krylov D, and Bergman D, 2000 Phys. Rev. Lett. {\bf 84} 1902 
\bibitem{ref.hatano} Hatano N, 2005 J. Phys. Soc. Jpn. {\bf 74} 3093
\bibitem{ref.redner} Krapivsky P L  and Redner S, 2004 Am. J. Phys. {\bf 72} 591
\bibitem{ref.esaki} Esaki K, Sato M, Kohmoto M, 2009 Phys. Rev. E {\bf 79}, 056226 
\bibitem{ref.peitgen} Peitgen H-O, Juergens H, and Saupe D, 2004 {\it Chaos and Fractals: New Frontiers of Science}
(Springer Verlag, New York) 
\bibitem{ref.ziff} Ziff R M and  McGrady E D, 1986 Macromolecules {\bf 19} 2513
\bibitem{ref.krapivsky} Krapivsky P L, 1992 J. Stat. Phys. {\bf 69} 125
\bibitem{ref.krapivsky-naim} Krapivsky P L and Ben-Naim E, 1994 Phys. Lett. A {\bf 196} 168
\bibitem{ref.hassan} Hassan M K, 1997 Physical Review E {\bf 55} 5302
\bibitem{ref.cda} Hassan M K and Hassan M Z, 2009 Phys. Rev. E E {\bf 79} 021406 
%\bibitem{dubuc}Belair,J.,Dubuc,S.,(eds.) Fractal geometry and analysis, Kluwer Academic Publishers, Dordrecht,Holland,1991.
%\bibitem{schro} Schroder,M., Fratals, Chaos and Power Laws, W.H. Freeman and Co., New York,1991.
%\bibitem{falcon} K. Falconer, Techniques in Fractal Geometry, Wiley, New York, 1997
%\bibitem{mandel} Mandelbrot B., The fractal Geometry of Nature. New York: W.H. Freeman,1982
%\bibitem{peitgen & jurgens} Peitgen H, Jurgens H,saupe D Chaos and Fractals. Berlin: Springer-Verlag;1992 
%\bibitem{turner} Turner M, Blackledg J, Andrews P: Fractal Geometry in Digital Imaging. London: Academic Press 1998
%\bibitem{mandelt} Mandelbrot B, Les objects fractals: forme hasard et dimension, 1975
%\bibitem{vics} Vicsek, T., Fractal Growth Phenomena, World Scientific, London,1989
%\bibitem{aharony} Aharony,A. and Feder J(eds.)Fractal in Physics, Physica D 38(1989)
%\bibitem{dick} Dick Oliver, Fractal Vision(Sams,Indiana,USA 1992)
%\bibitem{feder} Feder J., Fractals(Plenum, New York 1998)
%\bibitem{hassan} M.K. Hassan, J Kurths Can randomnessalone tunethe fractal dimension? Physica A 315(2002)342-352
%\bibitem{hutch}Hutchinoson,J.,Fractals and self-similarity, Indian University Journal of Mathematics 30(1981)
%\bibitem{r.pynn}R.Pynn,A.Skjeltrop,eds.:Scaling Phenomena in Disordered systems(Plenm, Newyork in 1986)
%\bibitem{roger} Stevens Roger T. Fractal Programming in C. Redwood City,California: M and T Books,1990
%\bibitem{edger} Edger,G., Measures, Topology and Fractal Geometry, Springer-Verlag, New York 1990
%\bibitem{barnsley} Barnsley, Michael. Fractal Everywhere, San Diego: Academic Press in 1998
%\bibitem{deva}Devany, Robert. Chaos, Fractals, and Dynamics: Experiments in Modern Mathematics. New York : Addition-Wesley,1989.
%\bibitem{peitg}Peitgen, Heinz-Otto Richter, Piter H. The beauty of fractals; Images of complex Dynamical Systems. New York: Sjpringer-Verlag,1996
%\bibitem{hans}Hans Lauwerier, Fractals: Endlessly Repeated Geometrical Figures, Translated by Sophia Gill-Hoffstadt, Princeton University Press, Princeton NJ, 1991.
%\bibitem{mandelb} Mandelbrot B.B.,An introduction to multifractal distribution functions, in: Fluctuations and Pattern Formation, H.E. Stanley and N.Ostrowsky(eds.) Kluwer Academic, Dordrecht,1988.
%\bibitem{leland}Leland et. Al On the self-similar nature of ethernet traffic IEEE/ACM Transactions on Networking Volume2,Issue1(February1994)
%\bibitem{anis}Md. Anisur Rahman: Kotch Curve Revisited With Randomness Research Report.
%\bibitem{mcmul}C.T. McMullen, Hausdroff dimension and conformal dynamics \rm{2}. Geometrical finite rational maps, Comment. Math.Helv.75(2003),535-593.
%\bibitem{mcmullen}C.T McMullen, Hausdroff dimension and conformal dynamics \rm{3}. Computation of dimension, J.Math120(1998),691-721. 
%\bibitem{dubuc}Belair,J.,Dubuc,S.,(eds.) Fractal geometry and analysis, Kluwer Academic Publishers, Dordrecht,Holland,1991.
%\bibitem{stanley}H.E. Stanley, N. Ostrowsky,eds.: On growth and form: Fractal and Non-Fractal Patterns in Physics(Martinus Nijfhoff, Dordrecht1985)
%\bibitem{khassan}M.K. Hassan Physical Review E 55(1997)
%\bibitem{mkhassan}M.K. Hassan Physical Review E 54 (1996)
%\bibitem{dubuc}Belair,J.,Dubuc,S.,(eds.) Fractal geometry and analysis, Kluwer Academic Publishers, Dordrecht,Holland,1991.
%\bibitem{schro}Schroder,M., Fratals Chaos and Power Laws, W.H. Freeman and Co., New York,1991.

 

\end{thebibliography}
\end{document}